\documentclass[preprint,aps,prd,floatfix,superscriptaddress,%
preprintnumbers]{revtex4}
\usepackage{graphicx}

\def\be{\begin{equation}}
\def\ee{\end{equation}}
\def\bea{\begin{eqnarray}}
\def\eea{\end{eqnarray}}
\def\bean{\begin{eqnarray*}}
\def\eean{\end{eqnarray*}}
\def\bary{\begin{array}}
\def\eary{\end{array}}

\def\bit{\begin{itemize}}
\def\eit{\end{itemize}}
\def\bwt{\begin{widetext}}
\def\ewt{\end{widetext}}

\def\ol{\overline}

\def\ub{{\bar u}}
\def\cb{{\bar c}}

\def\db{{\bar d}}
\def\sb{{\bar s}}
\def\bb{{\bar b}}
\begin{document}

\preprint{\vbox{ \hbox{MADPH-04-1372; CLNS-04-1869}
    \hbox{hep-ph/0404073}
  }}
\title{CHARMLESS $B \to PP$ DECAYS USING FLAVOR SU(3) SYMMETRY
\footnote{To be submitted to Phys.~Rev.~D.}}

\author{Cheng-Wei Chiang}
\email[e-mail: ]{chengwei@pheno.physics.wisc.edu}
\affiliation{Department of Physics, University of Wisconsin, Madison, WI 53706}
\author{Michael Gronau}
\email[e-mail: ]{gronau@physics.technion.ac.il}
\affiliation{Department of Physics, Technion -- Israel Institute of Technology,
  Haifa 32000, Israel}
\author{Jonathan L.~Rosner \footnote{On leave from Enrico Fermi Institute and
Department of Physics, University of Chicago.}}
\email[e-mail: ]{rosner@hep.uchicago.edu}
\affiliation{Laboratory of Elementary Particle Physics, Cornell University,
Ithaca, NY 14850}
\author{Denis A.~Suprun}
\email[e-mail: ]{d-suprun@uchicago.edu}
\affiliation{Enrico Fermi Institute and Department of Physics,
University of Chicago, 5640 S. Ellis Avenue, Chicago, IL 60637}

\date{\today}

\begin{abstract}
  
  The decays of $B$ mesons to a pair of charmless pseudoscalar ($P$) mesons are
  analyzed within a framework of flavor SU(3).  Symmetry breaking is taken into
  account in tree ($T$) amplitudes through ratios of decay constants; exact
  SU(3) is assumed elsewhere.  Acceptable fits to $B \to \pi \pi$ and $B \to K
  \pi$ branching ratios and $CP$ asymmetries are obtained with tree,
  color-suppressed ($C$), penguin ($P$), and electroweak penguin ($P_{EW}$)
  amplitudes.  Crucial additional terms for describing processes involving
  $\eta$ and $\eta'$ include a large flavor-singlet penguin amplitude ($S$) as
  proposed earlier and a penguin amplitude $P_{tu}$ associated with
  intermediate $t$ and $u$ quarks.  For the $B^+ \to \pi^+ \eta'$ mode a term
  $S_{tu}$ associated with intermediate $t$ and $u$ quarks also may be needed.
  Values of the weak phase $\gamma$ are obtained consistent with an earlier
  analysis of $B \to VP$ decays, where $V$ denotes a vector meson, and with
  other analyses of CKM parameters.

\end{abstract}

\pacs{13.25.Hw, 14.40.Nd, 11.30.Er, 11.30.Hv}

\maketitle

\section{INTRODUCTION \label{sec:intro}}

A central objective of the study of $B$ meson decays is to help determine the
phases and magnitudes of Cabibbo-Kobayashi-Maskawa (CKM) matrix elements,
through the measurement of branching ratios and $CP$-violating observables.  It
is important to have accurate and self-consistent information on CKM matrix
elements if they are ever to be compared with fundamental theories predicting
them.  At present no such theories exist.  A further objective is to learn
about possible new physics at higher mass scales, affecting rare $B$ decays by
giving observables that appear to be inconsistent with others.  One wishes to
know whether there are any sources of $CP$ violation other than the phases in
the CKM matrix first proposed by Kobayashi and Maskawa \cite{Kobayashi:fv}.

Charmless $B$ meson decays, many of whose branching ratios and $CP$ asymmetries
(CPA's) have been measured to good accuracy, are an interesting and useful set
of modes.  Following the method presented in Ref. \cite{Chiang:2003pm} for $B$
decays into a vector meson ($V$) and a pseudoscalar meson ($P$), we analyze
observables in $B$ decays into two pseudoscalar mesons ($B \to PP$ decays) in
the present paper.  From the results of fits involving a small set of invariant
amplitudes, one can extract information about the parameters in theory, compare
with other known constraints, and predict as-yet-unreported observables.  In
particular, the amplitudes contributing to two-body hadronic charmless $B$
decays involve only one nontrivial weak phase $\gamma$ within the standard
model (SM).  In a previous analysis of $B \to VP$ decays \cite{Chiang:2003pm},
we found good agreement between the favored range of $\gamma$ from a fit to the
$VP$ modes and that from fits to CKM parameters \cite{Hocker:2001xe} based on
other measurements.  It is therefore of great interest to see if the $PP$ modes
give a consistent result.

In the present analysis, we take flavor SU(3) symmetry \cite{DZ,SW,Chau,%
Gronau:1994rj,Gronau:1995hn,GR95,Grinstein:1996us} as a working hypothesis.
Motivated by factorization in tree-level amplitudes, we take symmetry breaking
due to decay constant differences into account in these amplitudes when
relating strangeness-conserving and strange-changing processes.
We leave the issue of SU(3) symmetry breaking in penguin-type amplitudes to
experimental data.  As a test, one can compare the $B^+ \to \pi^+ K^0$ mode
(involving purely a strangeness-changing QCD penguin amplitude) with the $B^+
\to K^+ \ol{K}^0$ and $B^0 \to K^0 \ol{K}^0$ modes (involving purely
strangeness-conserving QCD penguin amplitudes).  In the limit of flavor $SU(3)$
symmetry, they should differ by a ratio of CKM factors, $V_{cs}/V_{cd}$.  If
penguin amplitudes $P_{tu}$ associated with intermediate $t$ and $u$ quarks are
important, the predictions for these modes will be affected.

We find acceptable fits to $B \to \pi \pi$ and $B \to \pi K$ branching ratios
and $CP$ asymmetries with a combination of tree, color-suppressed ($C$),
penguin ($P$), and electroweak penguin ($P_{EW}$) amplitudes. In contrast to
an earlier analysis of $B \to PP$ decays \cite{Chiang:2003rb}, in order to
describe these decays we must introduce a rather large value of $|C/T|$ and a
non-trivial relative phase between $C$ and $T$.  A large $|C/T|$ value could
improve agreement between the QCD factorization approach and experiment
\cite{BN}.  Our conclusion is driven in part by the large branching ratio for
$B^0 \to \pi^0 \pi^0$ reported recently \cite{Aubert:2003hf,Chao:2003ue}.

The data on processes involving $\eta$ and $\eta'$ also have made some progress
since our earlier analysis \cite{Chiang:2003rb}.  Crucial additional terms for
describing these decays include not only a large flavor-singlet penguin
amplitude ($S$) as proposed (e.g.) in Refs.\ \cite{DGR}, but also a penguin
amplitude $P_{tu}$ associated with intermediate $t$ and $u$ quarks, and
(for the $B^+ \to \pi^+ \eta'$ mode) a term $S_{tu}$ associated with
intermediate $t$ and $u$ quarks.

Values of the weak phase $\gamma \simeq 60^\circ$ are obtained consistent with
our earlier analysis of $B \to VP$ decays \cite{Chiang:2003pm}.  Other robust
aspects of our fit include the magnitude of the strangeness-changing penguin
amplitude, the strong phase of the tree amplitude relative to the penguin
($\sim 20^\circ$--$30^\circ$), the size of electroweak penguin contributions,
the correlation of a large direct $CP$ asymmetry in $B^0 \to \pi^+ \pi^-$
with a small one in $B^0 \to \pi^- K^+$, the correct prediction of signs and
magnitudes of all other measured direct $CP$ asymmetries as well, and a
fairly large negative value of the time-dependent $CP$ asymmetry parameter
$S_{\pi \pi}$.  Some other aspects of the fit are less likely to remain
unchanged in the face of further data; we shall comment on them in due course.

We review our conventions for the quark content of pseudoscalar mesons and
topological amplitudes in Section \ref{sec:not}.  Experimental data and
topological decompositions of decay amplitudes are presented in Section
\ref{sec:amp}.  In Section \ref{sec:points} we enumerate the data that will be
used in our $\chi^2$ fit.  Two fits to $\pi \pi$ and $\pi K$ observables are
presented in Section \ref{sec:fit1}, while modes with $\eta$ or $\eta'$ in the
final state are included in Section \ref{sec:fit2}.  We comment on robust and
less-stable aspects of the fits in Section \ref{sec:cmts}.  Based upon our
fitting results, we discuss our predictions for as-yet-unreported modes in
Section \ref{sec:unseen}.  Comparisons with other recent approaches (e.g.,
Refs.\ \cite{BN,Keum:2003qi,Buras:2004ub,Zenczykowski:2004tw}) are pursued in
Section \ref{sec:comp}. We summarize our findings in Section \ref{sec:summary}.

\section{NOTATION \label{sec:not}}

Our quark content and phase conventions \cite{Gronau:1994rj,GR95} are:
\begin{itemize}
\item{ {\it Bottom mesons}: $B^0=d\bb$, ${\ol B}^0=b\db$, $B^+=u\bb$,
    $B^-=-b\ub$, $B_s=s\bb$, ${\ol B}_s=b\sb$;}
\item{ {\it Charmed mesons}: $D^0=-c\ub$, ${\ol D}^0=u\cb$, $D^+=c\db$,
    $D^-=d\cb$, $D_s^+=c\sb$, $D_s^-=s\cb$;}
\item{ {\it Pseudoscalar mesons}: $\pi^+=u\db$, $\pi^0=(d\db-u\ub)/\sqrt{2}$,
    $\pi^-=-d\ub$, $K^+=u\sb$, $K^0=d\sb$, ${\ol K}^0=s\db$, $K^-=-s\ub$,
    $\eta=(s\sb-u\ub-d\db)/\sqrt{3}$,
    $\eta^{\prime}=(u\ub+d\db+2s\sb)/\sqrt{6}$;}
\end{itemize}
The $\eta$ and $\eta'$ correspond to octet-singlet mixtures
\be
\eta  = \eta_8 \cos \theta_0 - \eta_1 \sin \theta_0~,~~
\eta' = \eta_8 \sin \theta_0 + \eta_1 \cos \theta_0~~,
\ee
with $\theta_0 = \sin^{-1}(1/3) = 19.5^\circ$.

In the present approximation there are seven types of amplitudes: a
``tree'' contribution $t$; a ``color-suppressed'' contribution $c$; a
``penguin'' contribution $p$; a ``singlet penguin'' contribution $s$, in which
a color-singlet $q \bar q$ pair produced by two or more gluons or by a $Z$ or
$\gamma$ forms an $SU(3)$ singlet state; an ``exchange'' contribution $e$, an
``annihilation'' contribution $a$, and a ``penguin annihilation'' contribution
$pa$.  These amplitudes contain both the leading-order and electroweak penguin
contributions, and appear in the independent combinations
\be\bary{lll}
\label{eqn:dict}
t \equiv T + P_{\rm EW}^C ~, &\quad& c \equiv C + P_{\rm EW} ~, \\
p \equiv P - P_{tu} - \frac{1}{3} P_{\rm EW}^C ~, &\quad&
s \equiv S - S_{tu} - \frac{1}{3} P_{\rm EW} ~, \\
a \equiv A ~, &\quad& e + pa \equiv E + PA ~,
\eary\ee
where the capital letters denote the leading-order contributions
\cite{Gronau:1994rj,GR95,Gronau:1995hn,DGR} while $P_{\rm EW}$ and $P_{\rm
  EW}^C$ are respectively color-favored and color-suppressed electroweak
penguin amplitudes \cite{Gronau:1995hn}.  We shall neglect smaller terms
\cite{EWVP,GR2001} $PE_{\rm EW}$ and $PA_{\rm EW}$ [$(\gamma,Z)$-exchange and
$(\gamma,Z)$-direct-channel electroweak penguin amplitudes].  We shall denote
$\Delta S = 0$ transitions by unprimed quantities and $|\Delta S| = 1$
transitions by primed quantities.  The hierarchy of these amplitudes can be
found in Ref.~\cite{Chiang:2001ir}.  By writing QCD and flavor-singlet penguins
as $P - P_{tu}$ and $S - S_{tu}$, we adopt the so-called $c$-quark convention,
in which the heavy top quark is integrated out from the theory.  For
penguin-type amplitudes, we use the unitarity relation $V_{tb}^* V_{td} +
V_{cb}^* V_{cd} + V_{ub}^* V_{ud}=0$ to remove any top quark dependence. The
$V_{ub}^* V_{ud}$ term of the top quark mediated penguin is combined with the
up quark mediated penguin to form $P_{tu}$ or $S_{tu}$.  Similarly, the
$V_{cb}^* V_{cd}$ term is united with the charm quark mediated penguin into $P$
or $S$.  As a consequence, the strangeness-conserving $P$ and $S$ and
strangeness-changing $P'$ and $S'$ penguin amplitudes have real weak phases in
our discussions.  The relation between the $c$-quark convention and the
$t$-quark convention, where the $c$ quark dependence is removed instead, can be
found in, e.g., Ref.\ \cite{conv}.

The partial decay width of two-body $B$ decays is
\be
\label{eq:width}
\Gamma(B \to M_1 M_2)
= \frac{p_c}{8 \pi m_B^2} |{\cal A}(B \to M_1 M_2)|^2 ~,
\ee
where $p_c$ is the momentum of the final state meson in the rest frame of $B$,
$m_B$ is the $B$ meson mass, and $M_1$ and $M_2$ can be either pseudoscalar or
vector mesons.  Using Eq.~(\ref{eq:width}), one can extract the invariant
amplitude of each decay mode from its experimentally measured branching ratio.
To relate partial widths to branching ratios, we use the world-average
lifetimes $\tau^+ = (1.653\pm0.014)$ ps and $\tau^0 = (1.534\pm0.013)$ ps
computed by the LEPBOSC group \cite{LEPBOSC}.  Unless otherwise indicated, for
each branching ratio quoted we imply the average of a process and its
$CP$-conjugate.

\section{AMPLITUDE DECOMPOSITIONS AND EXPERIMENTAL RATES \label{sec:amp}}

The experimental branching ratios and $CP$ asymmetries on which our analysis is
based are listed in Tables \ref{tab:dS0data} and \ref{tab:dS1data}.
Contributions from the CLEO \cite{Bornheim:2003bv,Richichi:1999kj,%
Chen:2000hv,Behrens:1998dn}, BaBar \cite{Aubert:2003qj, Aubert:2003sg,%
Aubert:2003xb,Aubert:2002jb,BaBarPlot0054,Aubert:2003hf, Aubert:2003bq,%
FryLP03,Aubert:2004xf}, and Belle \cite{Chao:2003ue,Belle0311,Huang:2002ev,%
Abe:2001pf,Abe:2004us,Unno:2003,Aihara,Chen:2002af,FryLP03,Tomura,Abe:2003yt}
Collaborations are included \cite{HFAG}.  In order to implement upper bounds in
a consistent manner we have computed our own experimental averages for $B^+ \to
\pi^+ \eta'$ and $B^0 \to \eta K^0$. These two modes were observed by BaBar
with a significance of 3.4 and 3.3 standard deviations, respectively.

\begin{table*}
\scriptsize
\caption{Experimental branching ratios of selected $\Delta S = 0$ decays of 
$B$ mesons. $CP$-averaged branching ratios are quoted in units of $10^{-6}$. 
Numbers in parentheses are upper bounds at 90 \% c.l.  References are given
in square brackets.  Additional lines, if any, give the $CP$ asymmetry 
${\cal  A}_{CP}$ (second line) or $({\cal S},{\cal A})$ (second and third 
lines) for charged or neutral modes, respectively.  The error in the average
includes the scale factor $S$ when this number is shown in parentheses.
\label{tab:dS0data}}
\begin{ruledtabular}
\begin{tabular}{llllll}
 & Mode & CLEO & BaBar & Belle & Average \\ 
\hline
$B^+ \to$
    & $\pi^+ \pi^0$ 
        & $4.6^{+1.8+0.6}_{-1.6-0.7}$ \cite{Bornheim:2003bv}
        & $5.5^{+1.0}_{-0.9}\pm0.6$ \cite{Aubert:2003qj}
        & $5.0\pm1.2\pm0.5$ \cite{Chao:2003ue}
        & $5.2\pm0.8$ \\
    &   & -
        & $-0.03^{+0.18}_{-0.17}\pm0.02$ \cite{Aubert:2003qj}
        & $-0.14\pm0.24^{+0.05}_{-0.04}$ \cite{Belle0311}
        & $-0.07\pm0.14$ \\
    & $K^+ \ol{K}^0$ 
        & $<3.3$ \cite{Bornheim:2003bv}
        & $1.1\pm0.75^{+0.14}_{-0.18} \; (<2.5)$ \cite{Aubert:2003sg}
        & $<3.3$ \cite{Chao:2003ue}
        & $<2.5$ \\
    & $\pi^+ \eta$ 
        & $1.2^{+2.8}_{-1.2} \; (<5.7)$ \cite{Richichi:1999kj}
        & $5.3\pm1.0\pm0.3$ \cite{Aubert:2003xb}
        & $5.4^{+2.0}_{-1.7} \pm 0.6$ \cite{Huang:2002ev}
        & $4.9\pm0.9$ \\  
    &   & - 
        & $-0.44\pm0.18\pm0.01$  \cite{Aubert:2003xb}
        & -
        & $-0.44\pm0.18$ \\
    & $\pi^+ \eta'$ 
        & $1.0^{+5.8}_{-1.0} \; (<12)$ \cite{Richichi:1999kj}
        & $2.7\pm1.2\pm0.3 \; (<4.5)$ \cite{Aubert:2003xb}
        & $<7$ \cite{Abe:2001pf}
        & $2.4 \pm 1.1~(<4.5)$ \\
\hline
$B^0 \to$
    & $\pi^+ \pi^-$ 
        & $4.5^{+1.4+0.5}_{-1.2-0.4}$ \cite{Bornheim:2003bv}
        & $4.7\pm0.6\pm0.2$ \cite{Aubert:2002jb}
        & $4.4\pm0.6\pm0.3$ \cite{Chao:2003ue}
        & $4.6\pm0.4$ \\
    &   & -
        & $\left\{\bary{l}
          -0.40\pm0.22\pm0.03 \\
          0.19\pm0.19\pm0.05
          \eary\right.$ \cite{BaBarPlot0054}
        & $\left\{\bary{l}
          -1.00\pm0.21\pm0.07 \\
          0.58\pm0.15\pm0.07
          \eary\right.$ \cite{Abe:2004us}
        & $\left\{\bary{l}
          -0.70\pm0.30 \, (S=1.91) \\
          0.42\pm0.19 \, (S=1.52)
          \eary\right.$ \\
    & $\pi^0 \pi^0$
        & $<4.4$ \cite{Bornheim:2003bv}
        & $2.1\pm0.6\pm0.3$ \cite{Aubert:2003hf} 
        & $1.7\pm0.6\pm0.2$ \cite{Chao:2003ue}
        & $1.9\pm0.5$ \\
    & $K^+ K^-$ 
        & $<0.8$ \cite{Bornheim:2003bv}
        & $<0.6$ \cite{Aubert:2002jb}
        & $<0.7$ \cite{Chao:2003ue}
        & $<0.6$ \\
    & $K^0 \ol{K}^0$ 
        & $<3.3$ \cite{Bornheim:2003bv}
        & $0.6^{+0.7}_{-0.5}\pm0.1 \; (<1.8)$ \cite{Aubert:2003sg}
        & $<1.5$ \cite{Chao:2003ue}
        & $<1.5$ \\
    & $\pi^0 \eta$ 
        & $0.0^{+0.8}_{-0.0} \; (<2.9)$ \cite{Richichi:1999kj}
        & $0.7^{+1.1}_{-0.9}\pm0.3 \; (<2.5)$ \cite{BabarULs}
        & -
        & $<2.5$ \\
    & $\pi^0 \eta'$ 
        & $0.0^{+1.8}_{-0.0} \; (<5.7)$ \cite{Richichi:1999kj} 
        & $1.0^{+1.4}_{-1.0}\pm0.8 \; (<3.7)$ \cite{BabarULs}
        & -
        & $<3.7$ \\
    & $\eta \eta$ 
        & $<18$ \cite{Behrens:1998dn}
        & $-0.9^{+1.6}_{-1.4}\pm0.7 \; (<2.8)$ \cite{Smith} & -
        & $<2.8$ \\
    & $\eta \eta'$ 
        & $<27$ \cite{Behrens:1998dn}
        & $0.6^{+2.1}_{-1.7}\pm1.1 \; (<4.6)$ \cite{Smith} & -
        & $<4.6$ \\
    & $\eta' \eta'$ 
        & $<47$ \cite{Behrens:1998dn}
        & $1.7^{+4.8}_{-3.7}\pm0.6 \; (<10)$ \cite{Smith} & -
        & $<10$ \\
\end{tabular}
\end{ruledtabular}
\end{table*}

\begin{table*}
\scriptsize
\caption{Same as Table \ref{tab:dS0data} for $|\Delta S| = 1$ decays of $B$
 mesons.
\label{tab:dS1data}}
\begin{ruledtabular}
\begin{tabular}{llllll}
 & Mode & CLEO & BaBar & Belle & Average \\ 
\hline
$B^+ \to$
    & $\pi^+ K^0$ 
        & $18.8^{+3.7+2.1}_{-3.3-1.8}$ \cite{Bornheim:2003bv}
        & $22.3\pm1.7\pm1.1$ \cite{Aubert:2003sg}
        & $22.0\pm1.9\pm1.1$ \cite{Chao:2003ue}
        & $21.8\pm1.4$ \\
    &   & $0.18\pm0.24\pm0.02$ \cite{Chen:2000hv}
        & $-0.05\pm0.08\pm0.01$ \cite{Aubert:2003sg}
        & $0.07^{+0.09+0.01}_{-0.08-0.03}$ \cite{Unno:2003}
        & $0.02\pm0.06$ \\
    & $\pi^0 K^+$ 
        & $12.9^{+2.4+1.2}_{-2.2-1.1}$ \cite{Bornheim:2003bv}
        & $12.8^{+1.2}_{-1.1}\pm1.0$ \cite{Aubert:2003qj}
        & $12.0\pm1.3^{+1.3}_{-0.9}$ \cite{Chao:2003ue}
        & $12.5\pm1.0$ \\
    &   & $-0.29\pm0.23\pm0.02$ \cite{Chen:2000hv}
        & $-0.09\pm0.09\pm0.01$ \cite{Aubert:2003qj}
        & $0.23\pm0.11^{+0.01}_{-0.04}$ \cite{Belle0311}
        & $0.00\pm0.12 \, (S=1.79)$ \\
    & $\eta K^+$ 
        & $2.2^{+2.8}_{-2.2} \; (<6.9)$ \cite{Richichi:1999kj}
        & $3.4\pm0.8\pm0.2$ \cite{Aubert:2003xb}
        & $5.3^{+1.8}_{-1.5}\pm0.6$ \cite{Huang:2002ev} 
        & $3.7\pm0.7$ \\
    &   & -
        & $-0.52\pm0.24\pm0.01$ \cite{Aubert:2003xb}
        & -
        & $-0.52\pm0.24$ \\
    & $\eta' K^+$ 
        & $80^{+10}_{-9}\pm7$ \cite{Richichi:1999kj}
        & $76.9\pm3.5\pm4.4$ \cite{Aubert:2003bq}
        & $78\pm6\pm9$ \cite{Aihara}
        & $77.6\pm4.6$ \\
    &   & $0.03\pm0.12\pm0.02$ \cite{Chen:2000hv}
        & $0.037\pm0.045\pm0.011$ \cite{Aubert:2003bq}
        & $-0.015\pm0.070\pm0.009$ \cite{Chen:2002af}
        & $0.02\pm0.04$ \\
\hline
$B^0 \to$
    & $\pi^- K^+$ 
        & $18.0^{+2.3+1.2}_{-2.1-0.9}$ \cite{Bornheim:2003bv}
        & $17.9\pm0.9\pm0.7$ \cite{Aubert:2002jb}
        & $18.5\pm1.0\pm0.7$ \cite{Chao:2003ue}
        & $18.2\pm0.8$ \\
    &   & $-0.04\pm0.16\pm0.02$ \cite{Chen:2000hv}
        & $-0.107\pm0.041\pm0.013$ \cite{FryLP03}
        & $-0.088\pm0.035\pm0.018$ \cite{FryLP03}
        & $-0.09\pm0.03$ \\
    & $\pi^0 K^0$ 
        & $12.8^{+4.0+1.7}_{-3.3-1.4}$ \cite{Bornheim:2003bv}
        & $11.4\pm1.7\pm0.8$ \cite{Aubert:2003sg}
        & $11.7\pm2.3^{+1.2}_{-1.3}$ \cite{Chao:2003ue}
        & $11.7\pm1.4$ \\
&   & - 
        & $\left\{\bary{l}
          0.48^{+0.38}_{-0.47}\pm0.06 \\
          -0.40^{+0.28}_{-0.27}\pm0.09
          \eary\right.$ \cite{Aubert:2004xf}
        & 
                & $\left\{\bary{l}
          0.48\pm0.42  \\
          -0.40\pm0.29
          \eary\right.$ \\   
   & $\eta K^0$ 
        & $0.0^{+3.2}_{-0.0} \; (<9.3)$ \cite{Richichi:1999kj}
        & $2.9\pm1.0\pm0.2 \; (<5.2)$ \cite{Aubert:2003xb} 
        & $<12$ \cite{Tomura}
        & $2.5 \pm 1.0 \, (S=1.08)~(<5.2)$ \\
    & $\eta' K^0$ 
        & $89^{+18}_{-16}\pm9$ \cite{Richichi:1999kj}
        & $60.6\pm5.6\pm4.6$ \cite{Aubert:2003bq}
        & $68\pm10^{+9}_{-8}$ \cite{Aihara} 
        & $65.2\pm6.2 \, (S=1.03)$ \\
    &   & - 
        & $\left\{\bary{l}
          0.02\pm0.34\pm0.03 \\
          -0.10\pm0.22\pm0.04
          \eary\right.$ \cite{Aubert:2003bq}
        & $\left\{\bary{l}
          0.43\pm0.27\pm0.05 \\
          -0.01\pm0.16\pm0.04
          \eary\right.$ \cite{Abe:2003yt}
        & $\left\{\bary{l}
          0.27\pm0.21 \\
          -0.04\pm0.13
          \eary\right.$ \\
\end{tabular}
\end{ruledtabular}
\end{table*}

We list theoretical predictions and averaged experimental amplitudes for
charmless $B \to PP$ decays involving $\Delta S = 0$ transitions in Table
\ref{tab:dS0} and those involving $|\Delta S| = 1$ transitions in Table
\ref{tab:dS1}.  Theoretical predictions are shown in terms of topological
amplitudes $t$, $c$, $p$ and $s$ while $e$, $a$ and $pa$ contributions are
neglected.  They are expected to be suppressed by a factor of order $1/m_b$ 
relative to tree and penguin amplitudes \cite{Bauer:2004tj}.  A suppression
factor proportional to $f_B/m_b$ was suggested in~\cite{Gronau:1994rj,%
Gronau:1995hn}.  Future measurements of the $B^0 \to K^+ K^-$ decay mode which
only receives contributions from exchange and penguin annihilation diagrams
will test this suppression.

\section{$\chi^2$ fit and data points \label{sec:points}}

We define for $n$ experimental observables $X_i \pm \Delta X_i$ and the
corresponding theoretical predictions $X_i^{\rm th}$,
%

\begin{table*}
\caption{Summary of predicted contributions to $\Delta S = 0$ decays of $B$
  mesons to two pseudoscalars.  Amplitude magnitudes $|A_{\rm exp}|$ extracted
  from experiments are quoted in units of eV.
\label{tab:dS0}}
\begin{ruledtabular}
\begin{tabular}{llccc}
 & Mode & Amplitudes & $p_c$ (GeV) & $|A_{\rm exp}|^a$ \\ 
\hline
$B^+ \to$
    & $\pi^+ \pi^0$ & $-\frac{1}{\sqrt{2}}(t+c)$
        & $2.636$ & $23.4\pm1.7$ \\
    & $K^+ \ol{K}^0$ & $p$ & $2.593$ & $<16.4$ \\
    & $\pi^+ \eta$ & $-\frac{1}{\sqrt{3}}(t+c+2p+s)$ 
        & $2.609$ & $22.9\pm2.0$ \\
    & $\pi^+ \eta'$ & $\frac{1}{\sqrt{6}}(t+c+2p+4s)$ 
        & $2.551$ & $16.2\pm3.8 \; (<22.2)$ \\
\hline
$B^0 \to$
    & $\pi^+ \pi^-$ & $-(t+p)$
        & $2.636$ & $22.8\pm1.1$ \\
    & $\pi^0 \pi^0$ & $-\frac{1}{\sqrt{2}}(c-p)$ 
        & $2.636$ & $14.7\pm1.8$ \\
    & $K^+ K^-$ & $-(e+pa)$ 
        & $2.593$ & $<8.3$ \\
    & $K^0\ol{K}^0$ & $p$ 
        & $2.592$ & $<13.2$ \\
    & $\pi^0\eta$ & $-\frac{1}{\sqrt{6}}(2p+s)$ 
        & $2.610$ & $<17.0$ \\
    & $\pi^0\eta'$ & $\frac{1}{\sqrt{3}}(p+2s)$ 
        & $2.551$ & $<20.9$ \\
    & $\eta\eta$ & $\frac{\sqrt{2}}{3}(c+p+s)$ 
        & $2.582$ & $<18.1$ \\
    & $\eta\eta'$ & $-\frac{\sqrt{2}}{3}(c+p+\frac52s)$ 
        & $2.522$ & $<23.4$ \\
    & $\eta'\eta'$ & $\frac{1}{3\sqrt{2}}(c+p+4s)$ 
        & $2.460$ & $<35.0$ \\
\end{tabular}
\end{ruledtabular}
\leftline{$^a$ $|A_{\rm exp}|$ is defined by Eq.~(\ref{eq:width}) as an 
amplitude related to a $CP$-averaged branching ratio quoted in Table I.}
\end{table*}
%

\begin{table*}
\caption{Same as Table \ref{tab:dS0} for $|\Delta S| = 1$ decays of $B$ mesons.
\label{tab:dS1}}
\begin{ruledtabular}
\begin{tabular}{llccc}
 & Mode & Amplitudes & $p_c$ (GeV) & $|A_{\rm exp}|$ \\ 
\hline
$B^+ \to$
    & $\pi^+ K^0$ & $p'$ 
        & $2.614$ & $48.2\pm1.6$ \\
    & $\pi^0 K^+$ & $-\frac{1}{\sqrt{2}}(p'+t'+c')$ 
        & $2.615$ & $36.6\pm1.5$ \\
    & $\eta K^+$ & $-\frac{1}{\sqrt{3}}(s'+t'+c')$
        & $2.588$ & $19.9\pm1.9$ \\
    & $\eta' K^+$ & $\frac{1}{\sqrt{6}}(3p'+4s'+t'+c')$ 
        & $2.528$ & $92.5\pm2.7$ \\
\hline
$B^0 \to$
    & $\pi^- K^+$ & $-(p'+t')$
        & $2.615$ & $45.7\pm1.0$ \\
    & $\pi^0 K^0$ & $\frac{1}{\sqrt{2}}(p'-c')$ 
        & $2.614$ & $36.6\pm2.2$ \\
    & $\eta K^0$ & $-\frac{1}{\sqrt{3}}(s'+c')$
        & $2.587$ & $17.0\pm3.5 \; (<24.6)$ \\
    & $\eta' K^0$ & $\frac{1}{\sqrt{6}}(3p'+4s'+c')$
        & $2.528$ & $88.0\pm4.2$ \\
\end{tabular}
\end{ruledtabular}
\end{table*}

\be
\chi^2 =
\sum_{i=1}^n \left( \frac{X_i^{\rm th} - X_i}{\Delta X_i} \right)^2 ~.
\ee
The data points are the branching ratios and the $CP$ asymmetries.  We write
the corresponding theoretical predictions in terms of topological amplitudes
and extract their magnitudes, weak phases and strong phases by minimizing
$\chi^2$.

Tables \ref{tab:dS0data} and \ref{tab:dS1data} contain a total of 
26 data points,
including 9 observables from $\Delta S = 0$ decays and 17 from $|\Delta S| = 1$
decays.  The modes involving $\pi \pi$ and $\pi K$ consist of the following
15 pieces of data:

\begin{itemize}
\item The $\pi^+ \pi^0$ decay involving the $t$ and $c$ amplitudes provides two
  data points.  Since both amplitudes have the same weak phase except for a
  small contribution from EWP, no significant CPA is expected.
\item The $\pi^+ \pi^-$ decay involves the $t$ and $p$ amplitudes with different
  weak phases.  Time-dependent CPA's have been observed by both BaBar and Belle
  groups.  Thus, this mode provide three data points.
\item The $\pi^0 \pi^0$ decay involving the $c$ and $p$ amplitudes only provides
  one data point because no CPA has been measured yet.
\item The $\pi^+ K^0$ decay involving only the $p'$ amplitude provides two data
  points, although no significant CPA is expected.  This mode plays a 
  dominant role in
  constraining the magnitude of the $P'$ amplitude.
\item The $\pi^0 K^+$ decay involving the $p'$, $t'$, and $c'$ amplitudes
  provides two data points.
\item The $\pi^- K^+$ decay involving the $p'$ and $t'$ amplitudes provides two
  data points.
\item The $\pi^0 K^0$ decay involves the $p'$ and $c'$ amplitudes.
  Time-dependent CPA's have been reported by the BaBar group.  Thus, this mode
  provides three data points.
\end{itemize}

Successful SU(3) fits to modes with an $\eta$ or $\eta'$ in the final state
require amplitudes beyond those mandated by the $\pi \pi$ and $\pi K$ fits.
A common feature of these modes, for example, is that they involve a flavor
singlet amplitude $s$ or $s'$.  Moreover, uncertainties in $\eta$ and $\eta'$
wave functions and possible SU(3) breaking effects can affect such fits
\cite{BN}, so we list these 11 data points separately: 

\begin{itemize}
\item The $\pi^+ \eta$ mode involving the combination $t+c+2p+s$ provides two
  data points.
\item The $\pi^+ \eta'$ mode involving the combination $t+c+2p+4s$ provides
  one data point.
\item The $\eta K^+$ mode involving the combination $s'+t'+c'$ provides two
  data points.  Note that it does not contain $p'$; all three contributing
  amplitudes are comparable in size.  One generally expects significant CPA as
  a result of the interference between tree-level and penguin-loop diagrams.
\item The $\eta K^0$ mode involving the combination $s'+c'$ provides one
  data point.
\item The $\eta' K^+$ mode involving the combination $3p'+4s'+t'+c'$ provides
  two data points.
\item The $\eta' K^0$ mode provides three data points, including the
  $CP$-averaged branching ratio and time-dependent CPA's.
\end{itemize}

\section{$\chi^2$ fit to $\pi \pi$ and $\pi K$ modes \label{sec:fit1}}

To avoid complication from uncertainties in the flavor-singlet amplitudes,
wave functions of $\eta$ and $\eta'$, and associated SU(3) breaking effects, we
first fit the fifteen $\pi \pi$ and $\pi K$ data points.  A study restricted to
$B \to K \pi$ decays based on similar assumptions was carried out in Refs.\ 
\cite{Imbeault:2003it}.  Guided by the relative importance of
strangeness-conserving and strangeness-changing transitions, we choose $T$,
$C$, $P'$, and $P_{tu}$ as our parameters.

We further fix the strong phase convention to be
\bea
T &=& |T|\, e^{i (\delta_T + \gamma)} ~, \\
C &=& |C|\, e^{i (\delta_T + \delta_C +\gamma)} ~, \\
P_{tu} &=& |P_{tu}|\, e^{i (\delta_{P_{tu}} + \gamma)}  ~, \\
P' &=& -|P'| ~. 
\eea
The phase convention is such that zero strong phases of $T$, $C$ and $P_{tu}$
amplitudes correspond to these amplitudes having a phase of $\gamma$ with 
respect to the penguin-type amplitude $P$.  Note that $\delta_C$ is defined as
a relative strong phase between the $C$ and $T$ amplitudes.
The extra minus signs for $P'$ comes from the relative weak phase $\pi$ 
between $P'=(V_{cs}/V_{cd})\,P$ and $P$ amplitudes.

The expressions for the $T'$, $C'$, $P'_{tu}$, $P$, and $P_{EW}$ are
obtained from the above equations taking into account the following ratios
\bea
&&  \frac{T'}{T}  =  \frac{V_{us}}{V_{ud}} \, \frac{f_K}{f_\pi} = 
\frac{\lambda}{1-\lambda^2/2} \frac{f_K}{f_\pi} \simeq 0.281 ~, \\
&&  \frac{C'}{C}  =  \frac{P'_{tu}}{P_{tu}} =  \frac{V_{us}}{V_{ud}} =
\frac{\lambda}{1-\lambda^2/2} \simeq 0.230 ~, \\
&&  \frac{P}{P'}  =  \frac{P_{EW}}{P'_{EW}}  = \frac{V_{cd}}{V_{cs}} =
-\, \frac{\lambda}{1-\lambda^2/2} \simeq -0.230 ~,
\eea
where $\lambda=0.224$~\cite{Battaglia:2003in}.  Therefore, a major SU(3)
breaking effect from the decay constant difference is included for tree-type
diagrams.  No such effect is considered for penguin-type
amplitudes because we do not expect factorization to work in such cases.
The ratio $P_{EW}/P'_{EW}=V_{cd}/V_{cs}$ is being used for
the simplicity of our analysis. We checked that using 
$P_{EW}/P'_{EW}=V_{td}/V_{ts}$ 
(to express t-quark dominance of EWP amplitudes) 
does not affect the results in any significant way.

We explore two approaches to fitting $\pi\pi$ and $\pi K$ data points. 
One of them (Fit II) uses Eqs.~(\ref{eqn:dict}) for the 
topological amplitudes $t$, $c$ and $p$:
\bea
t & \equiv & T + P_{\rm EW}^C ~, \\
c & \equiv & C + P_{\rm EW} ~, \\
p & \equiv & P - P_{tu} - \frac{1}{3} P_{\rm EW}^C ~.
\eea
Using these three equations, we can write the amplitude for 
any $\pi \pi$ or $\pi K$ decay mode 
in Tables \ref{tab:dS0} and \ref{tab:dS1} 
in terms of 9 parameters: weak phase $\gamma$, topological 
amplitudes $|T|$, $|C|$, $|P_{tu}|$, and $|P'|$, strong phases $\delta_T$, 
$\delta_C$, and $\delta_{P_{tu}}$, and a parameter $\delta_{EW}$. The latter
relates EW penguins to tree-level diagrams and will be defined below. 

In the other approach (Fit I) 
we use the fact that $P_{tu}$ has the same weak factors as tree-level 
amplitudes $T$ and $C$. This allows us to absorb the $P_{tu}$ penguin 
into redefined $\tilde{T}$ and $\tilde{C}$ amplitudes:
\bea
\tilde{T} \equiv T - P_{tu} ~, \\
\tilde{C} \equiv C + P_{tu} ~. 
\eea
By writing topological amplitudes $t$, $c$ and $p$ in terms of $\tilde{T}$,
$\tilde{C}$ and $P$ as
\bea
t = \tilde{T} + P_{\rm EW}^C ~, \\
c = \tilde{C} + P_{\rm EW} ~, \\
p = P - \frac{1}{3} P_{\rm EW}^C ~, 
\eea
we still get the correct expressions for $B \to \pi\pi$ and $B \to \pi K$ decay
amplitudes, except for $B^+ \to \pi^+ K^0$ and $B^+ \to \pi^0 K^+$.  In these
two cases the resulting expressions differ from the correct ones by a $P'_{tu}$
term. Compared to the dominant QCD penguin $P'$, this term is expected to be
small. Thus, Fit I gives a good description of $B \to \pi\pi$ and $B \to \pi K$
modes in terms of redefined tree-level amplitudes. The advantage of this
approach is a smaller number of fit parameters as both $|P_{tu}|$ and its
strong phase $\delta_{P_{tu}}$ are absorbed into $\tilde{T}$ and $\tilde{C}$.
Just 7 fit parameters are used in Fit I: weak phase $\gamma$, amplitudes
$|\tilde{T}|$, $|\tilde{C}|$, and $|P'|$, strong phases $\delta_{\tilde{T}}$
and $\delta_{\tilde{C}}$, and the $\delta_{EW}$ parameter.

A relation between the EWP amplitudes and the tree-type diagrams has been
found in Ref.\ \cite{NR} using Fierz transformation to relate EWP operators
with tree-level operators.  Explicitly, we have the relations
\be
\label{eqn:nr}
P'_{EW} = - \delta_{EW}\, T' e^{-i \gamma} 
= - \delta_{EW}\, |T'|\, e^{i\, \delta_T}    ~, 
\ee
\be
\label{eqn:nr2}
P'^C_{EW} = - \delta_{EW}\, C' e^{-i \gamma} 
= - \delta_{EW}\, |C'|\, e^{i (\delta_T + \delta_C)}   ~, 
\ee
where both the color-allowed and color-suppressed EWP amplitudes have
approximately the same proportionality constant
\be
\label{eqn:nr3}
\delta_{EW} \simeq 0.65 \pm 0.15 ~.
\ee
These relations determine both the magnitudes and phases of the EW penguins.
Their weak phases are equal to the weak phase of $P'$, i.e.\ to $-\pi$.  They
appear as the minus signs in Eqs.~(\ref{eqn:nr}) and~(\ref{eqn:nr2}).  We do
not use Eq.~(\ref{eqn:nr3}) as a constraint in our fit, but simply use
$\delta_{EW}$ as a fit parameter and check whether it comes out within the
expected bounds.

Eqs.~(\ref{eqn:nr}) and~(\ref{eqn:nr2}) were incorporated into Fit II 
but Fit I only employs
redefined $\tilde{T'}$ and $\tilde{C'}$ that cannot be directly related to
the EW penguins. Instead, we write
\be
P'_{EW}+P'^C_{EW} = - \delta_{EW}\, (T'+C')\, e^{-i \gamma}
= - \delta_{EW}\, (\tilde{T'}+\tilde{C'})\, e^{-i \gamma} 
\ee
and then neglect $P'^C_{EW}$ which is expected to be the smaller of the two
to obtain
\be
P'_{EW} \simeq - \delta_{EW}\, (\tilde{T'}+\tilde{C'})\, e^{-i \gamma} 
= - \delta_{EW}\, (|\tilde{T'}| e^{i\, \delta_T} 
+ |\tilde{C'}| e^{i (\delta_T + \delta_C)})  ~.
\ee
This relation for $P'_{EW}$ was used in Fit I while $P'^C_{EW}$ was set
to zero. 

The fitting parameters of both fits are shown in the columns for
Fit I and Fit II in Table \ref{tab:para}. 
An unusually large $|\tilde{C}/\tilde{T}| \approx 1.4$ ratio predicted 
by Fit I is an indication of large $|P_{tu}|$~\cite{Buras:2004ub}, 
destructive interference between $T$ and $P_{tu}$
contributions to the redefined tree amplitude $\tilde{T}$, and 
constructive interference between $C$ and $P_{tu}$
contributions to $\tilde{C}$. Indeed, Fit II which separates $P_{tu}$ and
tree-level amplitudes predicts $|P_{tu}|=14.9$ and a much more reasonable 
$|C/T| = 0.46$ ratio.

Fits I and II represent a completely satisfactory description of $B \to \pi
\pi$ and $B \to \pi K$ decay modes. The branching ratio for $B^0 \to \pi^0 K^0$ 
is predicted to be about $1.7 \sigma$ below the observed value, while that
for $B^+ \to \pi^0 K^+$ is predicted to be about $1.1 \sigma$ below
experiment.  These deviations could be hints of new physics
\cite{Buras:2004ub,Grossman:2003qi}, or simply due to underestimates of
neutral-pion detection efficiencies \cite{Gronau:2003kj}.  The predictions are
shown in the columns for Fits I and II in Tables \ref{tab:preds0} and
\ref{tab:preds1}.  Uncertainties for all predictions have been estimated
by scanning the parameter space and studying the parameter sets that
led to $\chi^2$ values no more than 1 unit above the minimum. The spread
in predictions corresponding to those parameter sets has determined the
uncertainties in predictions. The same method was used in an earlier analysis
of $B \to VP$ decays~\cite{Chiang:2003pm}.

\begin{figure}[t]
\includegraphics[width=.484\textwidth]{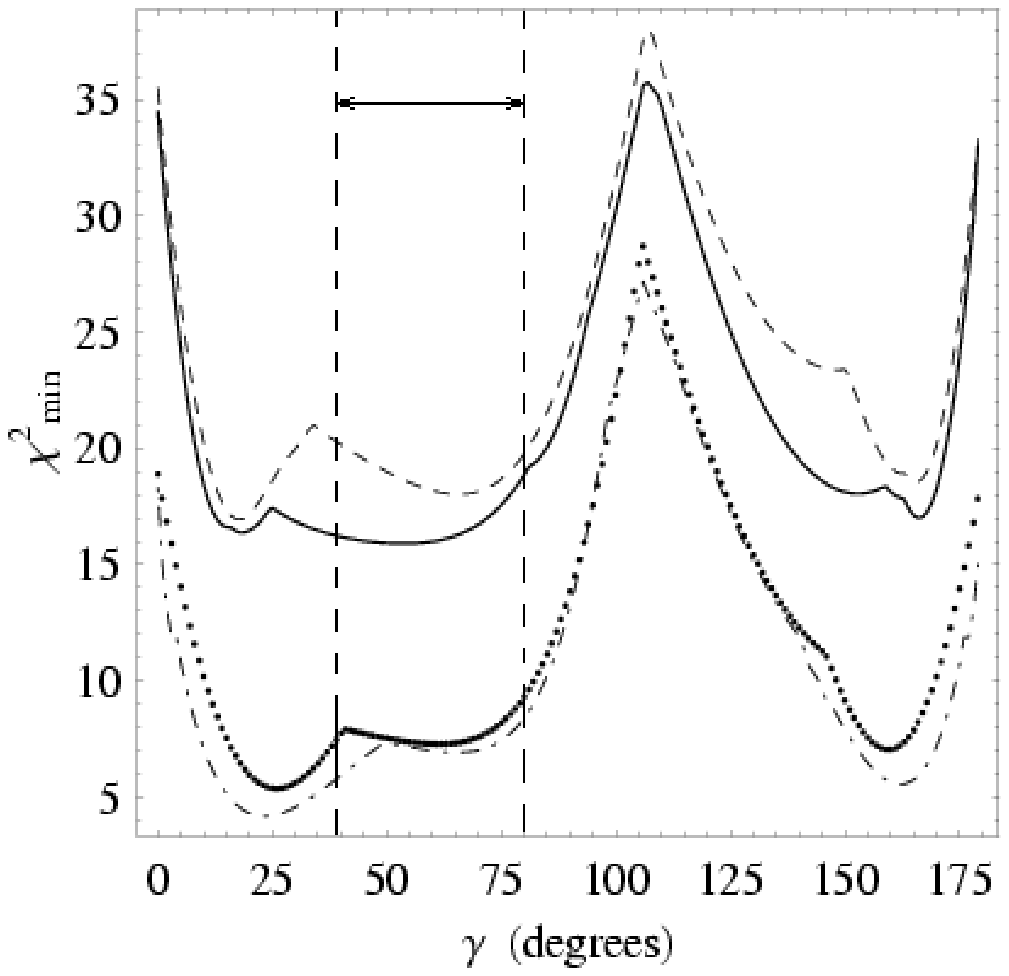}
\caption{$(\chi^2)_{\rm min}$, obtained by minimizing over all remaining fit
parameters, as a function of the weak phase $\gamma$.  Dotted curve: Fit I;
dash-dotted curve: Fit II; dashed curve: Fit III; solid curve: Fit IV.
Vertical dashed lines show the boundaries of the favored 95\% confidence level
range of $\gamma$ ($39^{\circ} - 80^{\circ}$)
from fits to CKM parameters~\cite{Hocker:2001xe} based on other measurements.
\label{fig:fit}}
\end{figure}

The confidence level of Fit II is slightly lower than in
Fit I because two new parameters ($|P_{tu}|$ and its strong phase
$\delta_{P_{tu}}$) have been added without a corresponding improvement in the
$\chi^2$ value.  The dependence of $\chi^2$ on the weak phase $\gamma$ in Fits
I and II is shown as the dotted and dash-dotted curves, respectively,
in Fig.~\ref{fig:fit}. 

\begin{table*}
\caption{Comparison of parameters extracted in fits to branching ratios and
$CP$ asymmetries under various assumptions. Values of the topological
amplitudes are quoted in units of eV.  The Fit I column shows values for
$\tilde{T}$ and $\tilde{C}$ and their strong phases in place of $T$ and $C$
amplitudes and phases.  Probabilities are those for $\chi^2$ to exceed the
value shown for the indicated number of degrees of freedom.
\label{tab:para}}
\small
\begin{ruledtabular}
\begin{tabular}{ccccc} 
Quantity & \multicolumn{2}{c}{Fit to $\pi\pi$,$\pi K$} &
 \multicolumn{2}{c}{Global fit} \\
 & Fit I & Fit II & Fit III & Fit IV \\
\hline
$\gamma$ & $(61^{+14}_{-27})^{\circ}$ & $(65^{+13}_{-35})^{\circ}$ 
& $(66^{+12}_{-16})^{\circ}$ & $(54^{+18}_{-24})^{\circ}$ \\
\hline
$|T|$ & $16.1^{+2.0}_{-1.9}$ & $30.4^{+15.1}_{-8.2}$ & $27.5\pm3.2$
& $27.4^{+7.9}_{-4.6}$ \\
$\delta_T$ & $(34^{+25}_{-11})^{\circ}$ & $(17^{+23}_{-12})^{\circ}$ 
& $(25 \pm 9)^{\circ}$ & $(34^{+17}_{-12})^{\circ}$ \\
$|C|$ & $22.9^{+4.3}_{-3.4}$ & $13.9^{+9.0}_{-8.5}$ & $19.2^{+3.1}_{-3.4}$
& $24.3^{+6.9}_{-5.1}$ \\
$\delta_C$ & $(-69^{+19}_{-22})^{\circ}$ & $(-94^{+43}_{-52})^{\circ}$
& $(-94^{+12}_{-11})^{\circ}$ & $(-103^{+17}_{-21})^{\circ}$ \\
$|P'|$ & $48.2^{+0.9}_{-1.0}$ & $47.7^{+0.8}_{-0.9}$ & $47.7 \pm 0.9$
& $47.8^{+0.9}_{-1.1}$ \\
$|P_{tu}|$ & 0 (input) & $14.9^{+14.0}_{-7.7}$ & $11.2 \pm 3.4$
& $12.3^{+7.7}_{-5.2}$ \\
$\delta_{P_{tu}}$ & 0 (input) & $(3^{+28}_{-27})^{\circ}$ & 
$(21 \pm 16)^{\circ}$ & $(37^{+17}_{-18})^{\circ}$ \\
$|S'|$ & 0 (input) & 0 (input) & $32.1^{+3.0}_{-3.3}$ 
& $32.4^{+2.9}_{-3.2}$ \\
$\delta_S$ & 0 (input) & 0 (input) & $(-69^{+11}_{-8})^{\circ}$
& $(-70^{+10}_{-8})^{\circ}$ \\
$|S_{tu}|$ & 0 (input) & 0 (input) & 0 (input)  & $5.7^{+5.5}_{-4.1}$ \\
$\delta_{S_{tu}}$ & 0 (input) & 0 (input) & 0 (input)
& $(-61^{+56}_{-42})^{\circ}$ \\
$\delta_{EW}$ & $0.55^{+0.44}_{-0.33}$ & $0.42^{+0.50}_{-0.29}$
& $0.47^{+0.32}_{-0.30}$ & $0.62^{+0.39}_{-0.36}$ \\
\hline
Fit properties: \\
$\chi^2 /$d.o.f. & 7.34/8 & 6.97/6 & 18.06/15 & 15.95/13 \\
CL (\%) & 50 & 32 & 26 & 25 \\
\hline
Derived quantities: \\
$|P'_{EW}|$ & $4.5^{+3.2}_{-2.6}$ & $3.6^{+3.6}_{-2.3}$ & $3.6^{+2.5}_{-2.3}$
& $4.8^{+4.3}_{-2.9}$ \\
$|P'^C_{EW}|$ & 0 (input) & $1.3^{+3.1}_{-1.0}$ & $2.1^{+1.6}_{-1.4}$ 
& $3.4^{+3.2}_{-2.2}$ \\
$|C/T|$ & $1.43^{+0.40}_{-0.31}$ & $0.46^{+0.43}_{-0.30}$ & $0.70 \pm 0.16$
& $0.89 \pm 0.21$ \\
\end{tabular}
\end{ruledtabular}
\end{table*}
\newpage

\begin{table*}
\caption{Comparison of predicted and experimental 
branching ratios in units of $10^{-6}$
and $CP$ asymmetries for $\Delta S = 0$ $B \to PP$ decays. The predictions
of Fits I and II for $\eta$ and $\eta'$ modes are not reliable and are given
for comparison purposes only.
$CP$ asymmetries, when predicted, are displayed
on second line for a decay mode, while asymmetries in curly brackets (when
shown) correspond to $S$ (second line) and $A$ (third line).
\label{tab:preds0}}
\small
\begin{ruledtabular}
\begin{tabular}{llccccc}
 & Mode & \multicolumn{2}{c}{Fit to $\pi \pi,\pi K$} &
   \multicolumn{2}{c}{Global fit} & Experimental \\
 & & Fit I & Fit II & Fit III & Fit IV & average \\
\hline
$B^+ \to$
    & $\pi^+ \pi^0$
        & $5.12^{+0.38}_{-0.23}$
        & $5.11^{+0.22}_{-0.14}$
        & $5.11^{+0.33}_{-0.37}$
        & $5.13^{+0.23}_{-0.22}$
        & $5.2\pm0.8$ \\
    &   & $-0.00 \pm 0.00$
        & $-0.00 \pm 0.00$
        & $-0.00 \pm 0.00$
        & $-0.01 \pm 0.00$
        & $-0.07\pm0.14$ \\
    & $K^+ \bar{K^0}$
        & $1.14 \pm 0.04$
        & $1.92^{+5.45}_{-1.35}$
        & $1.39^{+0.45}_{-0.35}$
        & $1.31^{+0.99}_{-0.36}$
        & $<2.5$ \\
    & $\pi^+ \eta$
        & $7.10^{+1.45}_{-1.05}$
        & $1.84^{+1.89}_{-0.39}$
        & $4.09^{+0.47}_{-0.41}$
        & $4.58^{+0.39}_{-0.51}$
        & $4.9\pm0.9$ \\
    &   & $-0.07^{+0.08}_{-0.06}$
        & $-0.40^{+0.90}_{-0.21}$
        & $-0.39^{+0.12}_{-0.11}$
        & $-0.40^{+0.09}_{-0.03}$
        & $-0.44\pm0.18$ \\
    & $\pi^+ \eta'$
        & $3.35^{+0.60}_{-0.46}$
        & $0.84^{+0.92}_{-0.19}$
        & $4.22^{+0.34}_{-0.31}$
        & $2.95^{+0.89}_{-0.55}$
        & $2.4\pm1.1 \ (<4.5)$ \\
    &   & $-0.07^{+0.08}_{-0.06}$
        & $-0.41^{+0.93}_{-0.21}$
        & $-0.10 \pm 0.10$
        & $-0.03^{+0.51}_{-0.34}$
        & $-$ \\
\hline
$B^0 \to$
    & $\pi^+ \pi^-$
        & $4.58^{+0.23}_{-0.28}$
        & $4.55^{+0.07}_{-0.06}$
        & $4.58^{+0.10}_{-0.12}$
        & $4.58^{+0.08}_{-0.11}$
        & $4.6\pm0.4$ \\
    &   & $\left\{\bary{l}
          -0.79^{+0.25}_{-0.16} \\
          0.34^{+0.02}_{-0.07}
          \eary\right.$
        & $\left\{\bary{l}
          -0.74^{+0.26}_{-0.21} \\
          0.33 \pm 0.02
          \eary\right.$
        & $\left\{\bary{l}
          -0.74^{+0.22}_{-0.16} \\
          0.31 \pm 0.06
          \eary\right.$
        & $\left\{\bary{l}
          -0.89^{+0.24}_{-0.06} \\
          0.30^{+0.02}_{-0.04}
          \eary\right.$
        & $\left\{\bary{l}
          -0.70\pm0.30  \\
          0.42\pm0.19 
          \eary\right.$ \\
    & $\pi^0 \pi^0$
        & $1.95^{+0.17}_{-0.30}$
        & $1.94^{+0.10}_{-0.18}$
        & $1.97^{+0.25}_{-0.27}$
        & $1.97^{+0.14}_{-0.19}$
        & $1.9\pm0.5$ \\
    &   & $\left\{\bary{l}
          0.44^{+0.35}_{-1.02} \\
          0.52^{+0.07}_{-0.20}
          \eary\right.$
        & $\left\{\bary{l}
          0.57^{+0.25}_{-1.30} \\
          0.53^{+0.03}_{-0.30}
          \eary\right.$
        & $\left\{\bary{l}
          0.54^{+0.22}_{-0.55} \\
          0.56^{+0.08}_{-0.10}
          \eary\right.$
        & $\left\{\bary{l}
          0.12^{+0.53}_{-0.83} \\
          0.52^{+0.09}_{-0.24}
          \eary\right.$
        & $\left\{\bary{l}
          -  \\
          - 
          \eary\right.$ \\
    & $K^0 \bar{K^0}$
        & $1.06 \pm 0.04$
        & $1.78^{+5.06}_{-1.25}$
        & $1.29^{+0.42}_{-0.32}$
        & $1.21^{+0.92}_{-0.33}$
        & $<1.5$ \\
    & $\pi^0 \eta$
        & $0.69 \pm 0.02$
        & $1.19^{+3.40}_{-0.83}$
        & $1.10^{+0.30}_{-0.33}$
        & $0.95^{+0.39}_{-0.16}$
        & $<2.5$ \\
    & $\pi^0 \eta'$
        & $0.31^{+0.02}_{-0.03}$
        & $0.57^{+1.67}_{-0.39}$
        & $1.34 \pm 0.18$
        & $1.00^{+0.49}_{-0.41}$
        & $<3.7$ \\
    & $\eta \eta$
        & $1.67^{+0.86}_{-0.44}$
        & $0.68^{+2.66}_{-0.47}$
        & $1.54^{+0.40}_{-0.29}$
        & $1.92^{+1.29}_{-0.48}$
        & $<2.8$ \\
    & $\eta \eta'$
        & $1.59^{+0.80}_{-0.42}$
        & $0.66^{+2.61}_{-0.47}$
        & $2.51^{+0.51}_{-0.36}$
        & $2.16^{+0.87}_{-0.60}$
        & $<4.6$ \\
    & $\eta' \eta'$
        & $0.38^{+0.18}_{-0.10}$
        & $0.16^{+0.64}_{-0.12}$
        & $0.97^{+0.16}_{-0.11}$
        & $0.68 \pm 0.32$
        & $<10$ \\
\end{tabular}
\end{ruledtabular}
\end{table*}
\newpage

\begin{table*}
\caption{Comparison of predicted and experimental
branching ratios in units of $10^{-6}$
and $CP$ asymmetries for $|\Delta S| = 1$ $B \to PP$ decays. The predictions
of Fits I and II for $\eta$ and $\eta'$ modes are not reliable and are given
for comparison purposes only.
$CP$ asymmetries, when predicted, are displayed
on second line for a decay mode, while asymmetries in curly brackets (when
shown) correspond to $S$ (second line) and $A$ (third line).
\label{tab:preds1}}
\small
\begin{ruledtabular}
\begin{tabular}{llccccc}
 & Mode & \multicolumn{2}{c}{Fit to $\pi \pi,\pi K$} &
   \multicolumn{2}{c}{Global fit} & Experimental \\
 & & Fit I & Fit II & Fit III & Fit IV & average \\ \hline
$B^+ \to$
    & $\pi^+ K^0$
        & $21.78^{+0.81}_{-0.82}$
        & $22.64^{+0.83}_{-0.93}$
        & $22.05^{+0.89}_{-0.95}$
        & $22.30^{+0.84}_{-0.78}$
        & $21.8\pm1.4$ \\
    &   & $0$
        & $0.00 \pm 0.04$
        & $0.03^{+0.02}_{-0.03}$
        & $0.05^{+0.02}_{-0.03}$
        & $0.02\pm0.06$ \\
    & $\pi^0 K^+$
        & $11.40^{+0.45}_{-0.70}$
        & $11.40^{+0.27}_{-0.72}$
        & $11.40^{+0.70}_{-0.77}$
        & $11.35^{+0.61}_{-0.68}$
        & $12.5\pm1.0$ \\
    &   & $0.02^{+0.03}_{-0.04}$
        & $0.03^{+0.05}_{-0.07}$
        & $0.07 \pm 0.02$
        & $0.09^{+0.01}_{-0.03}$
        & $0.00\pm0.12$ \\
    & $\eta K^+$
        & $0.16^{+0.04}_{-0.09}$
        & $0.21^{+0.07}_{-0.13}$
        & $3.44^{+0.60}_{-0.50}$
        & $3.63 \pm 0.59$
        & $3.7\pm0.7$ \\
    &   & $0$
        & $-0.10^{+0.08}_{-0.34}$
        & $-0.41^{+0.06}_{-0.04}$
        & $-0.34^{+0.11}_{-0.07}$
        & $-0.52\pm0.24$ \\
    & $\eta' K^+$
        & $29.38^{+0.58}_{-1.21}$
        & $30.72^{+1.06}_{-1.11}$
        & $74.56^{+1.51}_{-1.92}$
        & $75.21^{+1.44}_{-1.73}$
        & $77.6\pm4.6$ \\
    &   & $0.01 \pm 0.01$
        & $0.01^{+0.04}_{-0.05}$
        & $0.02 \pm 0.01$
        & $0.01 \pm 0.02$
        & $0.02\pm0.04$ \\
\hline
$B^0 \to$
    & $\pi^- K^+$
        & $18.90^{+0.46}_{-0.41}$
        & $18.60^{+0.50}_{-0.47}$
        & $18.89^{+0.45}_{-0.44}$
        & $18.78^{+0.48}_{-0.39}$
        & $18.2\pm0.8$ \\
    &   & $-0.10^{+0.02}_{-0.01}$
        & $-0.10 \pm 0.01$
        & $-0.10 \pm 0.02$
        & $-0.10^{+0.02}_{-0.01}$
        & $-0.09\pm0.03$ \\
    & $\pi^0 K^0$
        & $9.23^{+0.67}_{-0.47}$
        & $9.29^{+0.67}_{-0.50}$
        & $9.23^{+0.76}_{-0.65}$
        & $9.32^{+0.64}_{-0.63}$
        & $11.7\pm1.4$ \\
&       & $\left\{\bary{l}
          0.83 \pm 0.01 \\
          -0.11^{+0.04}_{-0.01}
          \eary\right.$
        & $\left\{\bary{l}
          0.83^{+0.01}_{-0.02}  \\
          -0.11^{+0.06}_{-0.01}
          \eary\right.$
        & $\left\{\bary{l}
          0.83 \pm 0.01  \\
          -0.12^{+0.03}_{-0.02}
          \eary\right.$
        & $\left\{\bary{l}
          0.83 \pm 0.01  \\
          -0.11^{+0.05}_{-0.02}
          \eary\right.$
        & $\left\{\bary{l}
          0.48\pm0.42  \\
          -0.40\pm0.29
          \eary\right.$ \\
    & $\eta K^0$
        & $0.07^{+0.00}_{-0.01}$
        & $0.05^{+0.10}_{-0.03}$
        & $2.66^{+0.46}_{-0.37}$
        & $2.49^{+0.43}_{-0.61}$
        & $2.5\pm1.0 \ (<5.2) $ \\
    &   & $\left\{\bary{l}
          -0.59^{+0.70}_{-0.19} \\
          0.60^{+0.35}_{-0.33}
          \eary\right.$
        & $\left\{\bary{l}
          0.34^{+0.55}_{-0.53} \\
          0.85^{+0.15}_{-0.48}
          \eary\right.$
        & $\left\{\bary{l}
          0.53^{+0.04}_{-0.03} \\
          0.02^{+0.05}_{-0.04}
          \eary\right.$
        & $\left\{\bary{l}
          0.56^{+0.03}_{-0.02} \\
          0.03^{+0.05}_{-0.03}
          \eary\right.$
        & $\left\{\bary{l}
          - \\
          -
          \eary\right.$ \\
    & $\eta' K^0$
        & $27.93^{+0.66}_{-1.09}$
        & $29.88^{+1.58}_{-1.47}$
        & $69.29^{+1.45}_{-1.84}$
        & $69.27^{+1.49}_{-1.72}$
        & $65.2\pm6.2 $ \\
    &   & $\left\{\bary{l}
          0.70^{+0.01}_{-0.00} \\
          0.04^{+0.00}_{-0.02}
          \eary\right.$
        & $\left\{\bary{l}
          0.81^{+0.09}_{-0.05} \\
          0.04 \pm 0.06
          \eary\right.$
        & $\left\{\bary{l}
          0.74 \pm 0.01 \\
          0.07 \pm 0.02
          \eary\right.$
        & $\left\{\bary{l}
          0.75^{+0.00}_{-0.01} \\
          0.06^{+0.01}_{-0.02}
          \eary\right.$
        & $\left\{\bary{l}
          0.27\pm0.21 \\
          -0.04\pm0.13
          \eary\right.$ \\
\end{tabular}
\end{ruledtabular}
\end{table*}

\section{Inclusion of modes with $\eta$ and $\eta'$ \label{sec:fit2}}

To enlarge the fit and discussion to decays involving $\eta$ or $\eta'$ in the
final state, we include an additional singlet amplitude.  It is represented by
\be
S' = -|S'| e^{i \delta_S} ~,
\ee
which gives two more fitting parameters. The relation between $S$ and $S'$
is the same as the one between $P$ and $P'$:
\be
\frac{S}{S'} = \frac{V_{cd}}{V_{cs}} = -\, \frac{\lambda}{1-\lambda^2/2} 
\simeq -0.230 ~. \\
\ee

The importance of the $S'$ amplitude has
been discussed in Refs.\ \cite{DGR,Chiang:2003rb,Chiang:2001ir} mainly to
account for the large branching ratios of the $\eta' K$ modes.  Moreover, we
include the parameter $P_{tu}$ and its associated strong phase 
$\delta_{P_{tu}}$. The penguin contribution $P_{tu}$ is apparently required 
by our fits to decay modes involving $\eta$ and $\eta'$.  For instance, in
$B^+ \to \pi^+ \eta^{(\prime)}$ the $P_{tu}$ contribution is of the same order
as the other terms and cannot be neglected.  The results under these
assumptions are given in the column for Fit III in Table \ref{tab:para}.
The $\chi^2$ dependence on $\gamma$ is shown as the dashed curve in
Fig.~\ref{fig:fit}.

Finally, since there is no reason to exclude such a term, we include a
contribution from a singlet-penguin amplitude $S_{tu}$ associated with
intermediate $t$ and $u$ quarks, consisting of a parameter $|S_{tu}|$ and its
associated strong phase $\delta_{S_{tu}}$:
\bea
S_{tu} &=& |S_{tu}| e^{i (\delta_{S_{tu}} + \gamma)} ~, \\
\frac{S'_{tu}}{S_{tu}} &=& \frac{V_{us}}{V_{ud}} = 
\frac{\lambda}{1-\lambda^2/2} \simeq 0.230 ~. 
\eea
This exercise is denoted by Fit IV.  The sole improvement
with respect to Fit III is a better fit to the $B^+ \to \pi^+ \eta'$
branching ratio, as shown in Table \ref{tab:preds0}.
The tree amplitude $|T|$ extracted from both Fit III and Fit IV
is in agreement with the estimate obtained from a recent application of
factorization \cite{Luo:2003hn} to the spectrum in
$B \to \pi l \nu$~\cite{Athar:2003yg}, which yields $24.4\pm3.8$~eV.

Both Fit III and Fit IV represent a good description of $B \to PP$ decay modes,
including those with $\eta$ or $\eta'$ in the final state.  The only
problematic data points are the branching ratio for $B^0 \to \pi^0 K^0$ which
is predicted to be about $1.7 \sigma$ below the observed value and the 
mixing-induced asymmetry $S(\eta' K^0)$ with the prediction $(\simeq \sin 2
\beta)$ at about $2.2 \sigma$ above the experimental value.  The predictions
for all other observed $\eta$ and $\eta'$ modes reproduce experimental values
within their uncertainties.  The predictions for as-yet-unseen modes are
consistent with the current experimental upper limits on their branching
ratios.  The predictions are shown in the columns for Fits III and IV in Tables
\ref{tab:preds0} and \ref{tab:preds1}.  The dependence of $\chi^2$ on $\gamma$
in Fit IV is shown as the solid curve in Fig.~\ref{fig:fit}.

\section{STABLE and LESS-STABLE ASPECTS OF FIT \label{sec:cmts}}

\subsection{Robust aspects}

The value of the weak phase $\gamma$ obtained in $B \to PP$ data is consistent
with other determinations.  All versions of the fits have a local $\chi^2$
minimum in the range $48^\circ \le \gamma \le 73^\circ$ (68\% c.l.) allowed by
global fits to phases of the CKM matrix \cite{Hocker:2001xe} and near the range
$(63 \pm 6)^\circ$ obtained in a fit to $B \to VP$ data \cite{Chiang:2003pm}.
The variation of $\gamma$ from fit to fit is at most about 12 degrees,
providing some idea of the systematic error associated with this approach.

All fits are comfortable with a relatively large negative value of $S_{\pi
  \pi}$ which is the average of the Babar \cite{BaBarPlot0054} and Belle
\cite{Abe:2004us} values.  Large negative $S_{\pi \pi}$ is associated with
larger $\alpha$ and smaller $\gamma$ (see, e.g., the plots in Refs.\ 
\cite{GRpipi}).

The magnitude $|P'|$ of the strangeness-changing penguin amplitude changes
very little from fit to fit.  It is specified by the decay $B^+ \to \pi^+
K^0$, which is expected to receive no other significant contributions.  The
presence of any direct $CP$ violation in this decay would call that assumption
into question, but no such asymmetry has yet been detected.

All fits obtain a much larger value of $|C/T|$ than the range of 0.08 to 0.37
assumed in Ref.\ \cite{Chiang:2003rb}.  Moreover, all fits (including those to
$\pi \pi$ and $\pi K$ modes alone) entail a large strong relative phase
$\delta_C$ between the $C$ and $T$ amplitudes.  The presence of a large
color-suppressed amplitude is somewhat of a surprise from the standpoint of
{\it a priori} calculations such as those in the QCD factorization approach
\cite{BN}, and probably indicates a greater-than-anticipated role for
final-state rescattering, which can generate such an effective amplitude (see
also \cite{Barshay:2004ra}).  Such rescattering may be the reason why the decay
$B^0 \to \pi^0 \pi^0$ is more prominent than had been expected.  All our fits
now entail a branching ratio for this mode of about $2 \ \times 10^{-6}$.
Although the favored values of some topological amplitudes (e.g., $C$,
$P_{tu}$) show noticeable variations from fit to fit, they change together in a
correlated way so that the predictions for almost all of the modes that involve
them remain very stable.

The strong phase $\delta_T$ of the tree amplitude $T$ with respect to the
penguin amplitude $P$ is found to be non-zero and of the order of $20^\circ$ to
$30^\circ$.  It is most likely driven by the need to simultaneously describe
a large direct $CP$ asymmetry (the parameter $A_{\pi \pi}$) in $B^0 \to \pi^+
\pi^-$ and a small but significant direct asymmetry in $B^0 \to \pi^- K^+$.
These quantities are well-fitted and their predicted values do not differ
much among the four fits.

While the electroweak penguin parameter $\delta_{EW}$ was initially constrained
to lie within the range (\ref{eqn:nr3}), we found that leaving it as a free
parameter led to results consistent with that range except in the cases of Fit
II and Fit III.  Thus, our fits do not favor a large phenomenological EWP
amplitude.  This should be contrasted with
Refs.~\cite{Buras:2004ub,Yoshikawa:2003hb} where a different assignment of weak
and strong phases is given in expectation of new physics contributions.  Our
fits also do not favor much deviation of the predicted $S_{\pi^0 K^0}$
time-dependent asymmetry parameter from its predicted standard-model value of
$\sin(2 \beta) \simeq 0.74$ \cite{Gronau:2003kx}.

Once one admits enough parameters into the fits to correctly describe modes
involving $\eta$ and $\eta'$, the negative direct $CP$ asymmetry in $B^+ \to
\pi^+ \eta$ observed by BaBar \cite{Aubert:2003xb} is correctly reproduced.
The possibility that this asymmetry could be large was first noted in Ref.\ 
\cite{Barshay:1990gy} and pursued in Refs.\ \cite{DGR}.  We predict a similarly
large negative $CP$ asymmetry in $B^+ \to \eta K^+$, as observed
\cite{Aubert:2003xb}.  These asymmetries can be large because no single weak
amplitude dominates the decays.  As sensitivities of asymmetric $e^+ e^-$
collider experiments improve through the accumulation of larger data samples,
we expect more such decay modes to emerge.

The mixing-induced and direct asymmetries $S(\eta' K^0)$ and $A(\eta' K^0)$ are
predicted to be close to $\sin(2 \beta)$ and $0$, respectively. These two
values would be expected if the $B^0 \to \eta' K^0$ decay amplitude had
consisted of just QCD penguin $P'$ and singlet penguin $S'$. The interference
of these terms with the much smaller $C'$, $P'_{tu}$, and $S'_{tu}$ amplitudes
leads to small deviations from the expected values. These deviations are to a
large extent determined by the ratio $|A'_C/A'_P|$ of the terms with the weak
factor $V_{ub}^* V_{us}$ ($C'$, $P'_{tu}$, and $S'_{tu}$) and the terms with
the weak factor $V_{cb}^* V_{cs}$ ($P'$ and $S'$).  $|A'_C/A'_P|$ is typically 
predicted by QCD factorization and PQCD to be smaller than 0.02~\cite{BN,%
London:1997zk,Kou:2001pm}.  Our best conservative estimate of $|A'_C/A'_P|$ is
based on Fit IV.  We find that the SU(3) fit prefers somewhat larger values:
$|A'_C/A'_P|=0.042^{+0.017}_{-0.006}$.  Fit III (somewhat more stable than Fit
IV) predicts $|A'_C/A'_P|=0.040^{+0.011}_{-0.009}$.  More conservative bounds
on $|A'_C/A'_P|$ and on the asymmetries $S(\eta' K^0)$ and $A(\eta' K^0)$ were
obtained recently in a model-independent way using flavor SU(3)
\cite{GR:etaprK}.

We have explored the effects of changing the $\eta$--$\eta'$ octet-singlet
mixing angle from its nominal value $\theta \simeq 19.5^{\circ}$ defined in
Sec.\ II.  The angle $\theta$ assumed a value of $22.0^{\circ}$ in Fit III with
a free mixing angle while $\chi^2$ of the fit improved by just 1.12.  With one
additional parameter in the fit, this did not result in a better fit quality.
Fit IV with a free mixing angle preferred $\theta=20.4^{\circ}$, with the fit
quality dropping by 5\%.  Thus, leaving the $\eta$--$\eta'$ mixing angle as a
free parameter, we found variations of only a few degrees and negligible
improvements in fits.

\subsection{Aspects sensitive to assumptions}

The possibility of a large $P_{tu}$ term in Fit II leads to a wide range of
predicted branching ratios for $B^+ \to K^+ \bar{K^0}$ and
$B^0 \to K^0 \bar{K^0}$. This range is considerably reduced in other fits.

The magnitude and phase of the singlet penguin amplitude $S'$ are probably not
well-determined.  The two quantities are correlated, as first pointed out in
Ref.\ \cite{DGR} and noted further in Ref.\ \cite{Chiang:2003rb}.  For example,
a much smaller magnitude of $S'$ is required to fit the charged and neutral $B
\to \eta' K$ decay modes if $S'$ and $P'$ (the gluonic penguin amplitude)
interfere constructively with one another.  The QCD factorization approach
\cite{BN} finds negligible $S'$ contribution to these decays, explaining their
enhancement by means of nonet-symmetry breaking effects as proposed, for
example, in Ref.\ \cite{FKS}, and making use of the constructive interference
of non-strange and strange components of the $\eta'$ in the gluonic penguin
amplitude \cite{HJLP}.  One should also point out that many other explanations
have been proposed for the enhancement of $B \to \eta' K$ modes
\cite{Kou:2001pm,Keta}.  One also finds the magnitude of $S'$ to be sensitive
to small changes in the octet-singlet mixing in $\eta$ and $\eta'$.

Predictions for the branching ratio and $CP$ asymmetry in $B^+ \to \eta K^+$
depend crucially on the introduction of the $S'$ amplitude.  Since this
amplitude is uncertain in magnitude and phase, those predictions (although
apparently satisfied) should be viewed with caution.  The same warning
applies to the mode $B^+ \to \eta \pi^+$.

As already noted, the predicted branching ratio for $B^+ \to \pi^+ \eta'$ is
quite sensitive to assumptions, and was the sole quantity which could be
compared to experiment that led to the introduction of the $S_{tu}$ term in 
Fit IV.  In
Ref.\ \cite{Chiang:2003rb} we noted a tight correlation between predicted
branching ratios and $CP$ asymmetries for $B^+ \to \pi^+ \eta$ and $B^+ \to
\pi^+ \eta'$.  With the added possibility of nonzero $P_{tu}$ and $S_{tu}$
contributions, this correlation no longer holds.

The only other prediction whose values are significantly different in Fits III
and IV is the mixing-induced asymmetry $S(\pi^0 \pi^0)$. One should trust the
larger values of this quantity predicted by Fits I and II. These fits to $\pi
\pi$ and $\pi K$ data points are not affected by the uncertainties associated
with $\eta$ and $\eta'$. Their predictions for the asymmetries in $B^0 \to
\pi^0 \pi^0$ modes thus are expected to be more reliable.

The introduction of the $S_{tu}$ term changes the favored value of $\gamma$ by
a noticeable amount, though still within limits from CKM global fits
\cite{Hocker:2001xe}.  As noted, this provides one estimate of systematic
errors associated with analyses of the present form.

\section{MODES TO BE SEEN \label{sec:unseen}}

Several decay modes are predicted to occur at levels just below present upper
bounds, and can provide useful constraints on the residual uncertainties in our
fits.  For example, the decays $B^+ \to K^+ \ol{K}^0$ and $B^0 \to K^0
\ol{K}^0$ are predicted to have branching ratios exceeding $10^{-6}$ (somewhat
larger than in Ref.\ \cite{Chiang:2003rb}), with the exact value depending on
the fit.  The decay modes $B^0 \to \pi^0 \eta$ and $B^0 \to \pi^0 \eta'$ also
should be visible at this level.  The modes $B^0 \to (\eta \eta, \eta
\eta',\eta'\eta')$ will probably require more work.  We also make predictions
for the direct and mixing-induced asymmetries in $B^0 \to \pi^0 \pi^0$ and $B^0
\to \eta K^0$, with $A(\pi^0 \pi^0)$ exceeding 0.5. A prediction for the
branching ratio of $K^+K^-$ cannot be made in our approach. The amplitude of
this decay mode receives contributions from exchange and penguin annihilation
diagrams that are neglected in this paper. It is very desirable that a more
strict experimental upper limit be set for this mode to justify the assumption
of negligibility of similar contributions to other neutral $\Delta S=0$ decay
modes.

\begin{table}
\caption{Comparison of observed and predicted direct $CP$ asymmetries for
some $B \to \pi \pi$ and $B \to \pi K$ decay modes.  \label{tab:pikdir}}
\begin{ruledtabular}
\begin{tabular}{c c c c c c c}
Decay & Exptl.\ & Present  & \multicolumn{2}{c}{QCDF \cite{BN}}
 & PQCD \cite{Keum:2003qi} & Ref.\ \cite{Buras:2004ub} \\
Mode  & average & work (a) & Full range & Favored (b) & & \\ 
\hline
$B^0 \to \pi^+ \pi^-$ & $0.42 \pm 0.19$ & $0.30^{+0.02}_{-0.04}$
 & $-0.07^{+0.14}_{-0.13}$ & 0.10 & 0.16 to 0.30 & Input \\
$B^+ \to \pi^- K^+$ & $-0.09 \pm 0.03$ & $-0.10^{+0.02}_{-0.01}$
 & $0.04^{+0.09}_{-0.10}$ & $-0.04$ & $-(0.13~{\rm to}~0.22)$
 & $-0.14^{+0.09}_{-0.14}$ \\
$B^0 \to \pi^0 K^0$ & $-0.40 \pm 0.29$ & $-0.11^{+0.05}_{-0.02}$
 & $-0.03 \pm 0.04$ & 0.01 & -- & $-0.05^{+0.29}_{-0.24}$ \\ 
\end{tabular}
\end{ruledtabular}
\vskip 0.1in
\centerline{(a) Fit IV; (b) Scenario S4}
\end{table}

\begin{table}
\caption{Comparison of observed and predicted direct $CP$ asymmetries for
some $B$ decay modes involving $\eta$ and $\eta'$. \label{tab:etadir}}
\begin{ruledtabular}
\begin{tabular}{c c c c c} 
Decay & Exptl.\ & Present  & \multicolumn{2}{c}{QCDF \cite{BN}} \\
Mode  & average & work (a) & Full range & Favored (b) \\
\hline
$B^+ \to \pi^+ \eta$ & $-0.44 \pm 0.18$ & $-0.40^{+0.09}_{-0.03}$
 & $-0.15 \pm 0.20$ & 0.06 \\
$B^+ \to \eta K^+$ & $-0.52 \pm 0.24$ & $-0.34^{+0.11}_{-0.07}$
 & $-0.19 ^{+0.29}_{-0.30}$ & 0.10 \\ 
\end{tabular}
\end{ruledtabular}
\vskip 0.1in
\centerline{(a) Fit IV; (b) Scenario S4}
\end{table}

\section{COMPARISON WITH OTHER APPROACHES \label{sec:comp}}

The signs of our predicted direct $CP$ asymmetries agree with those measured
experimentally for the five processes in which non-zero asymmetries are
reported at greater than the $2 \sigma$ level.  We summarize these and our
predictions for them in Tables \ref{tab:pikdir} and \ref{tab:etadir}.  (For
others, as shown in Tables VI and VII, negligible asymmetries are predicted, in
accord with observation.)

The fact that we agree with all five signs and magnitudes is due in part to the
flexibility of our SU(3) fit, but still represents a non-trivial consistency in
our description of strong phases.  We were not able to achieve this consistency
in Ref.\ \cite{Chiang:2003rb}.  The same correlation between predicted signs of
direct asymmetries in $B^0 \to \pi^+ \pi^-$ and $B^0 \to \pi^- K^+$ occurs in
all the methods compared in Table \ref{tab:pikdir}.  A definite prediction of
the absolute signs, in accord with experiment, is made in Ref.\
\cite{Keum:2003qi}.

Fits to $B \to PP$ branching ratios in the various approaches which we compare
with ours \cite{BN,Keum:2003qi,Buras:2004ub,Zenczykowski:2004tw} are generally
acceptable, especially when allowance is made for possible large penguin
amplitudes and color-suppressed contributions.  These fits now are converging
on a preference for $\gamma$ in the range preferred by fits
\cite{Hocker:2001xe} to other observables constraining CKM parameters.

\section{SUMMARY \label{sec:summary}}

We have analyzed the decays of $B$ mesons to a pair of charmless pseudoscalar
mesons within a framework of flavor SU(3).  Acceptable fits to $B \to \pi \pi$
and $B \to K \pi$ branching ratios and $CP$ asymmetries were obtained with
tree, color-suppressed ($C$), penguin ($P$), and electroweak penguin ($P_{EW}$)
amplitudes, but in order to describe processes involving $\eta$ and $\eta'$ we
needed to include a large flavor-singlet penguin amplitude ($S$) and a penguin
amplitude $P_{tu}$ associated with intermediate $t$ and $u$ quarks.  For the
$B^+ \to \pi^+ \eta'$ mode a term $S_{tu}$ associated with intermediate $t$ and
$u$ quarks also was employed.

We were able to achieve a good fit to the five most significant direct $CP$
asymmetries, as noted in Tables \ref{tab:pikdir} and \ref{tab:etadir}.  We
found values of the weak phase $\gamma$ roughly consistent with those obtained
earlier in an analysis of $B \to VP$ decays [($\gamma = 63 \pm 6)^\circ$], and
with other analyses \cite{Hocker:2001xe} of CKM parameters, for which the 68\%
confidence level limit is $48^\circ \le \gamma \le 73^\circ$.  A global fit
without $S_{tu}$ gave $\gamma = (66^{+12}_{-16})^\circ$, 
while adding $S_{tu}$ yielded
$\gamma = (54^{+18}_{-24})^\circ$. The difference between these two 
serves as an estimate of systematic error.

\section*{ACKNOWLEDGMENTS}

We thank P.~Chang, C.~Dallapiccola, D.~London, S. Mishima, D.~Pirjol, A. I.
Sanda, J.~G.~Smith, and T. Yoshikawa for helpful discussions.  C.-W.~C thanks
the hospitality of the Particle Physics Theory Group at Cornell University
during his visit when part of this work was done.  J. L. R. thanks M. Tigner
for extending the hospitality of the Laboratory for Elementary-Particle Physics
at Cornell university during this investigation, and the John Simon Guggenheim
Memorial Foundation for partial support.  This work was supported in part by
the United States Department of Energy, High Energy Physics Division, through
Grant Nos.\ DE-FG02-90ER40560, DE-FG02-95ER40896, and W-31-109-ENG-38.

\end{document}